\documentclass[prd,twocolumn,superscriptaddress,floatfix,nofootinbib,10pt]{revtex4}
\usepackage{setspace}
\usepackage[utf8]{inputenc}
\usepackage{float}
\usepackage{amsmath,amssymb,amsfonts,bm}
\usepackage{graphicx}
\usepackage{multirow}
\usepackage[usenames,dvipsnames]{color}
\allowdisplaybreaks
\usepackage[
colorlinks=true,
linkcolor=blue,
breaklinks=true,
urlcolor=blue,
citecolor=blue]{hyperref}

\usepackage{orcidlink}
\usepackage{subfigure}
\usepackage{soul}
\usepackage{xcolor}
\usepackage{booktabs}
\usepackage{ulem}
\usepackage{makecell}
\definecolor{green}{rgb}{0,0.6,0}
\newcommand{\mev}{\textrm{ MeV}}

\newcommand{\PreserveBackslash}[1]{\let\temp=\\#1\let\\=\temp}
\newcolumntype{C}[1]{>{\PreserveBackslash\centering}p{#1}}
\newcolumntype{R}[1]{>{\PreserveBackslash\raggedleft}p{#1}}
\newcolumntype{L}[1]{>{\PreserveBackslash\raggedright}p{#1}}

\def \R{\textcolor{black}}

\begin{document}
	\title{Scattering data and correlation function for the $K f_1(1285)$ interaction}
	
	\begin{abstract}
	We study the interaction of a kaon with the $f_1(1285)$ resonance, assuming that the $f_1(1285)$ is a molecular state generated by the $K \bar K^*, \bar K K^*$ interaction, evaluating the scattering amplitude, the scattering length and effective range of the $K f_1$ system. The scattering amplitude develops a resonant structure approximately \R{$56$ MeV} below the $K f_1$ threshold, with a width of around \R{$123$ MeV} MeV.  The corresponding correlation function has the  distinctive shape of a system with a bound state close to threshold. We also show that the interaction of the $K f_1$ system  differs significantly from the one obtained assuming that the $f_1(1285)$ is an \R{ordinary, non-molecular,} particle. This provides motivation to continue the search for these observables, already initiated by the measurement of the $p f_1(1285)$ correlation function by the ALICE collaboration.

	\end{abstract}
	
\author{Wen-Hao Jia\orcidlink{0009-0001-1170-3540}}
\affiliation{Department of Physics, Guangxi Normal University, Guilin 541004, China}
\affiliation{Guangxi Key Laboratory of Nuclear Physics and Technology,
Guangxi Normal University, Guilin 541004, China}

\author{Jing Song\orcidlink{0000-0003-3789-7504}}
\email[]{Song-Jing@buaa.edu.cn}
\affiliation{Center for Theoretical Physics, School of Physics and Optoelectronic Engineering, Hainan University, Haikou 570228, China}

\author{Wei-Hong Liang\orcidlink{0000-0001-5847-2498}}%
\email[]{liangwh@gxnu.edu.cn}
\affiliation{Department of Physics, Guangxi Normal University, Guilin 541004, China}
\affiliation{Guangxi Key Laboratory of Nuclear Physics and Technology,
Guangxi Normal University, Guilin 541004, China}

\author{Eulogio Oset\orcidlink{0000-0002-4462-7919}}%
\email[]{oset@ific.uv.es}
\affiliation{Department of Physics, Guangxi Normal University, Guilin 541004, China}
\affiliation{Departamento de Física Teórica and IFIC, Centro Mixto Universidad de Valencia-CSIC Institutos de Investigación de Paterna, 46071 Valencia, Spain}	

\maketitle

\section{Introduction}\label{sec:Intr}
The use of correlation functions to learn about hadron dynamics has entered a new phase with the study of correlation functions for three particles \cite{DelGrande:2021mju,ALICE:2022boj,ALICE:2023gxp,ALICE:2023bny,Garrido:2024pwi}.
A particularly promising case is the measurement of the correlation functions of a particle with a resonance, which could correspond to a bound state of two elementary particles, or a particle and a clear bound state of the deuteron \cite{Rzesa:2024nra,Ramos:2025ibe}.
The first steps towards the particle-resonance correlation functions have been given recently in Refs.~\cite{Privatecomm,ALICE:2024rjz} by measuring the correlation function for the $pf_1(1285)$ system. 
\R{In the present work,} the $f_1(1285)$ is considered as a molecular state stemming from the $K \bar K^* - \bar K K^*$ interaction in isospin $I=0$ \cite{Xie:2019iwz,Lutz:2003fm,Roca:2005nm,Garcia-Recio:2010enl,Zhou:2014ila,Geng:2015yta,Lu:2016nlp}, and it has passed many tests supporting this picture: The decays to $a_0(980) \pi$ and $f_0(980) \pi$ and into $K\bar K \pi$ have been tested successfully in Refs.~\cite{Aceti:2015zva,Aceti:2015pma}.
Further consistency with this picture has also been found by studying different reactions: $K^- p \to f_1(1285) \Lambda$ in Ref.~\cite{Xie:2015wja}, $J/\psi \to \phi f_1(1285)$ in Ref.~\cite{Xie:2015lta}, $B^0_s \to J/\psi f_1(1285)$ in Ref.~\cite{Molina:2016pbg}, $\tau \to f_1(1285) \pi \nu_\tau$ in Ref.~\cite{Oset:2018zgc}, and $\bar B^0 \to J/\psi f_1(1285)$ in Ref.~\cite{He:2021exv}.
\R{There are other pictures for the  $f_1(1285)$,  mostly ordinary $q\bar q$ mesons~\cite{Godfrey:1985xj,Kirchbach:1995ep,Close:1997nm,Li:2000dy,Liu:2014doa,Volkov:2018fyv}, and even speculations that it could have a seizable glueball component~\cite{Moreira:2017ulo}. However, it is fair to say that no picture has undergone the thorough scrutiny that the molecular one has gone through, as mentioned above.}

The interaction of the proton with $f_1(1285)$ would be similar to the interaction of an external particle with a nucleus, in this case, a two-nucleon bound state, the deuteron.
Actually, the $K^- d$ interaction at low energies has been the subject of intense study \cite{Thomas:1979xu,Eisenberg:1980bf,Kamalov:2000iy,Ivanov:2004bv,Meissner:2005bz,Meissner:2006gx,Shevchenko:2011ce,Doring:2011xc,Mai:2014uma}.

The experimental efforts to measure the $p f_1(1285)$ correlation function stimulated theoretical work to make predictions prior to the experimental results in Ref.~\cite{Encarnacion:2025lyf}.
In this theoretical paper, the fixed center approximation (FCA) to the Faddeev equations \cite{Foldy:1945zz,Brueckner:1953zz,Brueckner:1953zza,Chand:1962ec,Barrett:1999cw,Deloff:1999gc,Kamalov:2000iy} was used, and reviews on the subject can be found in Refs.~\cite{MartinezTorres:2020hus,Roca:2010tf,Malabarba:2024hlv}.
A shortcoming of the FCA is that, in general, it does not satisfy  elastic unitarity at the threshold of the external particle and the cluster.
In Ref.~\cite{Encarnacion:2025lyf}, it was restored by multiplying the three-body amplitude by a factor close to unity, but the general problem was exposed. 
The solution to this problem was provided in Ref.~\cite{Ikeno:2025bsx}, in the study of the $n \bar D^*_{s0}(2317)$ interaction, establishing the relationship with the traditional way to deal with the interaction of particles with nuclei by means of an optical potential used with the Lippmann-Schwinger equation.
The framework of this modified FCA is very well suited to study the particle interaction with a composite state, since the FCA defines a cluster, in this case the composite state, with which the external particle interacts.
The framework in Ref.~\cite{Ikeno:2025bsx} was then applied to the study of $n \bar D_{s1}(2460)$ and $n \bar D_{s1}(2536)$ interactions \cite{Agatao:2025ckp}, likely to be measured in the future by the ALICE collaboration, where some simplified formulas were presented, and was also used to make predictions of a superexotic bound state $K^* D^* K^*$ in Ref.~\cite{Jia:2025obs}.

In the present work, taking advantage \R{of the fact} that the techniques to detect the $f_1(1285)$ (through its decay into $K \bar K \pi$) are already available in the ALICE collaboration, we present results for the $K f_1(1285)$ interaction, looking at threshold scattering parameters, the correlation function and the prediction of a bound state.
These observables could be extracted from the measured correlation function using the inverse method developed in Refs.~\cite{Ikeno:2023ojl,Albaladejo:2023wmv}, and they would serve as a test for the theoretical ideas concerning the compositeness of the $f_1(1285)$ resonance, the theoretical framework used and the existence of a bound state associated with this interaction.

\section{formalism}\label{sec:form}
\begin{figure}[t]
	\begin{center}
		\includegraphics[width=0.48\textwidth]{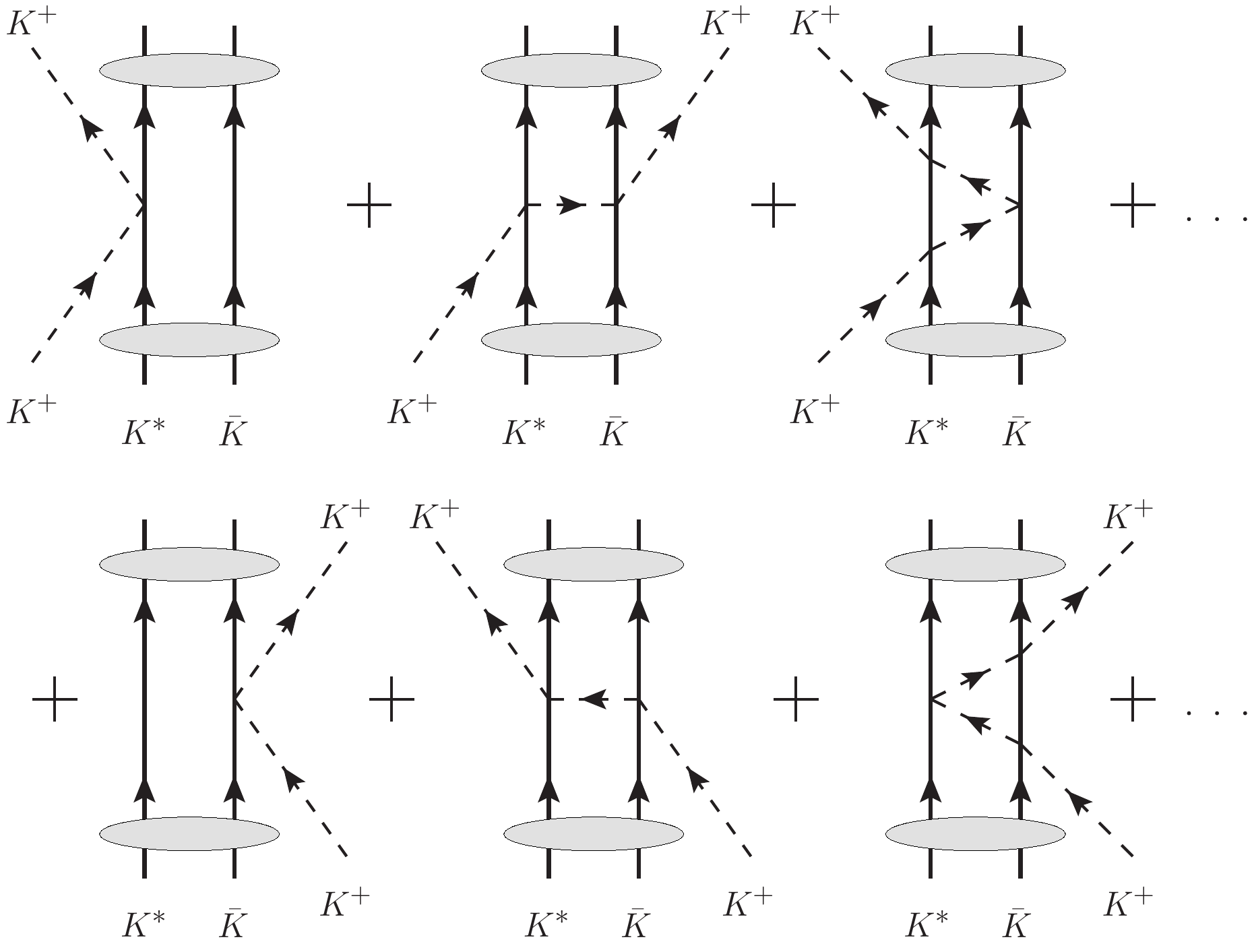}
	\end{center}
	\vspace{-0.5cm}
	\caption{Diagrams entering the ordinary FCA approach for $K^+$ interacting with the cluster of $K^*\bar K$.}
	\label{Fig1}
\end{figure}

We follow closely the formalism of Refs.~\cite{Ikeno:2025bsx,Agatao:2025ckp,Jia:2025obs}, in particular the one of Ref.~\cite{Jia:2025obs} which involves only mesons, given the different field normalization that we use for mesons and baryons following the nomenclature of Ref.~\cite{mandl_quantum_2010}.
We have two sums of diagrams, those of Fig.~\ref{Fig1}, which correspond to $K^+$ interacting with the $K^* \bar{K} (I=0)$ component of the $f_1(1285)$ state, considered in the conventional FCA approach.
We have to also consider the similar diagrams for the interaction of $K^+$ with the $\bar{K}^* K$ component.
For reasons of normalization to refer the amplitude to the $K^+ f_1(1285)$ system, we write the individual amplitudes
\begin{equation}
	\tilde{t}_1 = \frac{M_c}{M_{K^*}}\, t_1 \,\,,
	\qquad
	\tilde{t}_2 = \frac{M_c}{M_{K}}\, t_2 \,\,,
	\label{eq:1}
\end{equation}
where $M_c$ is the mass of the cluster $f_1(1285)$ \R{(we take $M_c=m_{f_1}=1281.8$~MeV)} and $t_1, t_2$ are the scattering matrices for the interaction of $K^+$ with $K^*$ and $\bar{K}$ respectively.
Considering that $K^* \bar{K}$ are in $I=0$ in the cluster we find that \cite{Ikeno:2025bsx}
\begin{equation}
	\begin{split}
		t_1 &= \frac{3}{4} \,t_{KK^*}^{I=1} + \frac{1}{4} \,t_{KK^*}^{I=0},\\[1.5mm]
		t_2 &= \frac{3}{4}\, t_{K\bar{K}}^{I=1} + \frac{1}{4} \,t_{K\bar{K}}^{I=0}.
	\end{split}
	\label{eq:2}
\end{equation}
Then we define $\tilde{T}_{ij}$ the sum of the diagrams where the $K^+$ interacts first with particle $i$ of the cluster and finishes in particle $j$ of the cluster in the sum of diagrams of Fig.~\ref{Fig1}.
One obtains
\begin{equation}
    \tilde{T} = \begin{pmatrix} \tilde{T}_{11} & \tilde{T}_{12} \\[1mm] 
	\tilde{T}_{21} & \tilde{T}_{22} \end{pmatrix},
\end{equation}
with
\begin{equation}
	\begin{split}
		& \tilde{T}_{11} = \frac{\tilde{t}_1}{1 - \tilde{t}_1 \,\tilde{t}_2 \,G_0^2}, ~~~~~~~~~~~~
		\tilde{T}_{22} = \frac{\tilde{t}_2}{1 - \tilde{t}_1\, \tilde{t}_2 \, G_0^2}, \\[1.5mm]
		& \tilde{T}_{12} = \tilde{T}_{21} = \frac{\tilde{t}_1 \,\tilde{t}_2 \,G_0}{1 - \tilde{t}_1 \,\tilde{t}_2 \, G_0^2},
	\end{split}
	\label{eq:3}
\end{equation}
where
\begin{align}\label{eq:4}
	G_0(\sqrt{s}) &= \int \frac{\mathrm{d}^3 q}{(2\pi)^3} \,
		\frac{F_c(q)}{\sqrt{s} - \omega_{K}(\vec q\,) - \omega_c(\vec q \,) + i \epsilon} \; \dfrac{1}{2\, \omega_{K}(\vec q\,)} \nonumber \\[1.5mm]
	&\times \dfrac{1}{2\, \omega_{c}(\vec q\,)}\; \Theta\left(q_{\mathrm{max}}^{(1)} - q_1^*\right) \, \Theta\left(q_{\mathrm{max}}^{(2)} - q_2^*\right),
\end{align}
with $\omega_{K}(\vec q\,) = \sqrt{M_{K^+}^2 + \vec{q}^{\;2}}$ and $\omega_c(\vec q\,) = \sqrt{M_c^2 + \vec{q}^{\;2}}$.
In Eq.~\eqref{eq:4}, $F_c(q)$ is the form factor of the cluster, which, considering the wave function in momentum space for the cluster coming from our approach \cite{Gamermann:2009uq,Yamagata-Sekihara:2010kpd}
{\color{black}
\begin{equation}
	\Psi(p)=g\;\frac{\Theta\left(q_{\mathrm{max}}-|\vec{p}\,|\right)}
	{M_c - \omega_{K^*}(\vec{p}\,) - \omega_{\bar K}(\vec p\,)},
	\label{eq:5}
\end{equation}
}
\noindent with $g$ the coupling of the state to the $K^* \bar{K}$ component, can be written as
\begin{equation}
	\begin{aligned}
		F_c(q) &= \frac{F(q)}{N}, \\[1.5mm]
		F(q) &= \int\limits_{\substack{|\vec{p}\,| < q_{\mathrm{max}} \\ |\vec{p} - \vec{q}\,| < q_{\mathrm{max}}}} 
		\frac{\mathrm{d}^3 p}{(2\pi)^3} \,
		\frac{1}{M_c - \omega_{K^*}(\vec{p}\,) - \omega_{\bar K}(\vec{p}\,)} \\
		&\quad \times \frac{1}{M_c - \omega_{K^*}(\vec{p} - \vec{q}\,) - 
			\omega_{\bar K}(\vec{p} -\vec{q}\,)}, \\[2mm]
		N &= F(0) \\
		&= \int\limits_{|\vec{p}\,| < q_{\mathrm{max}}} \frac{\mathrm{d}^3 p}{(2\pi)^3} \,
		\left[ \frac{1}{M_c - \omega_{K^*}(\vec{p}\,) - \omega_{\bar K}(\vec{p}\,)} \right]^2.
	\end{aligned}
	\label{eq:6}
\end{equation}
In Eq.~\eqref{eq:5}, $q_{\mathrm{max}}$ is the cutoff in the meson-meson loop functions, which arises naturally from assuming a potential of the type $V\left(\vec{q},\vec{q}\,'\right)=V\;\Theta\left(q_{\mathrm{max}}-|\vec{q}\,|\right)\;\Theta\left(q_{\mathrm{max}}-|\vec{q}\,'|\right)$, leading to the same factorization in the scattering matrix $t\left(\vec{q},\vec{q}\,'\right)=t\;\Theta\left(q_{\mathrm{max}}-|\vec{q}\,|\right)\;\Theta\left(q_{\mathrm{max}}-|\vec{q}\,'|\right)$. 
\R{We have $q_{\mathrm{max}}=1000$~MeV  from Ref.~\cite{Roca:2005nm} which predicts an accurate mass for the  $f_1(1285)$  (see Table~II in Appendix).}
In Eq.~\eqref{eq:4}, $q_{\mathrm{max}}^{(1)}$ refers to the cutoff associated with the $K^+ K^*$ scattering matrix and $q_{\mathrm{max}}^{(2)}$ to the $K^+ \bar{K}$ scattering matrix, 
\R{the values of which are found in the Appendix,}
and $q_1^*,\,q_2^*$ are the momenta of the $K^+$ in the rest frame of the $K^+K^*,\,K^+\bar K$, respectively, for the intermediate $K^+$, assuming the \R{momentum transferred to be shared equally between the initial and final particle of the cluster~\cite{Carrasco:1991we,Boffi:1991nh}}.
One has~\cite{Jia:2025obs}
\begin{equation}
	\vec{q}^{\,*}_i = \vec{q}\;\left( 1 - \frac{1}{2}\, \frac{M_{K^+}}{M_{K^+} + M_{i}} \right),
	\label{eq:7}
\end{equation}
with $M_1=M_{K^*}$, $M_2=M_{\bar K}$.

\begin{figure}[t]
	\begin{center}
		\includegraphics[width=0.48\textwidth]{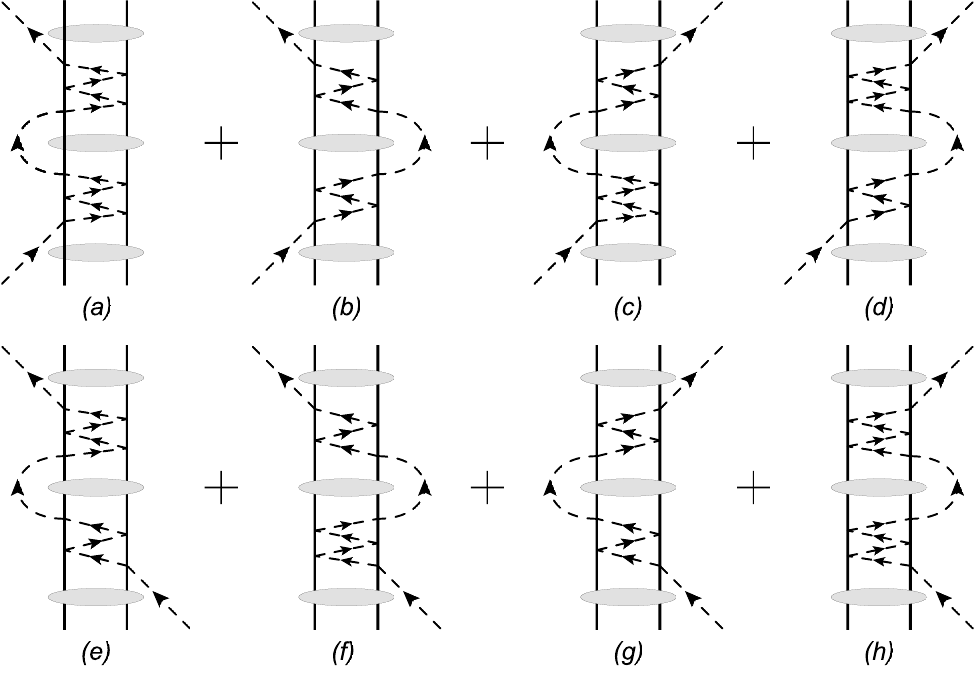}
	\end{center}
	\vspace{-0.5cm}
	\caption{Diagrams considering the elastic propagation of the $K^+$ and the cluster $f_1(1285)$ as a whole.}
	\label{Fig2}
\end{figure}

The unitarization in the elastic $K^+f_1(1285)$ channel is achieved by propagating the intermediate $K^+f_1(1285)$ state, which is equivalent to summing the diagrams of Fig.~\ref{Fig2}.
These sums lead to the terms
\begin{equation}
	\tilde{T}' = \begin{pmatrix} \tilde{T}_{11}' & \tilde{T}_{12}' \\[1mm]
	\tilde{T}_{21}' & \tilde{T}_{22}' \end{pmatrix}\,, \\
	\label{eq:8}
\end{equation}
where $\tilde{T}'_{ij}$ sums again all terms where the $K^+$ interacts first with particle $i$ of the cluster and finishes with particle $j$ of the cluster.
Considering further iterations of the $K^+f_1(1285)$ intermediate states in Fig.~\ref{Fig2}, one obtains
\begin{equation}\label{eq:9}
	\tilde{T}'=\left[ 1-\tilde{T}\,G_c \right]^{-1}\;\tilde{T},
\end{equation}
where
\begin{equation}
	G_c=\begin{pmatrix}
	G_c^{(1)} & 0 \\[1mm]
	 0 & G_c^{(2)}
	    \end{pmatrix},
\end{equation}
and
\begin{equation}\label{eq:10}
	\begin{aligned}[b]
		G_c^{(i)}(\sqrt{s}) &= \int \frac{\mathrm{d}^3 q}{(2\pi)^3} \;
		\frac{\left[ F_c^{(i)}(q) \right]^2}{\sqrt{s} - \omega_{K}(\vec q\,) - \omega_c(\vec q\,) + \mathrm{i}\epsilon} \\[1.5mm]
		&\quad \times \dfrac{1}{2\,\omega_{K}(\vec q\,)} \;\dfrac{1}{2\,\omega_{c}(\vec q\,)}\;\Theta\left(q_{\mathrm{max}}^{(i)} - q_i^*\right),
	\end{aligned}
\end{equation}
with \cite{Yamagata-Sekihara:2010kpd}
\begin{equation}
	\begin{aligned}
		F_c^{(1)}(q) &= F_c\left( \frac{M_{\bar K}}{M_{K^*}+M_{\bar K}}\, q\right),\\[2mm]
		F_c^{(2)}(q) &= F_c\left( \frac{M_{K^*}}{M_{K^*}+M_{\bar K}}\, q\right) \,.
	\end{aligned}
	\label{eq:11}
\end{equation}

One more ingredient is needed, the arguments of the $t_1$ and $t_2$ amplitudes, given by
\begin{equation}
	\begin{split}
		s(K^+ K^*) &= \left(p_{K^+} + p_{K^*}\right)^2 \\[1mm]
		&= M_{K^+}^2 + \left(\xi \, M_{K^*}\right)^2 + 2\,\xi\,M_{K^*}\,q^0,\\[2mm]
		s(K^+ \bar K) &= \left(p_{K^+} + p_{\bar K}\right)^2 \\[1mm]
		&= M_{K^+}^2 + \left(\xi \, M_{\bar K}\right)^2 + 2\,\xi\,M_{\bar K}\,q^0,
	\end{split}
	\label{eq:12}
\end{equation}
with $q^0$ the energy of the $K^+$ in the rest frame of the cluster
\begin{equation}\label{eq:13}
	 q^0=\frac{s-M_{K^+}^2-M_c^2}{2\,M_c},
\end{equation}
and
\begin{equation}\label{eq:14}
	\xi=\frac{M_c}{M_{K^*} + M_{\bar K}},
\end{equation}
which assumes that the binding of the $ f_1(1285)$ is shared between the $ K^*$ and the $ \bar K$ proportional to their masses.

The sum of the $ \tilde{T}'_{ij}$ matrices over $i,\,j$, which represents the total final amplitude for the $K^+f_1(1285)$ scattering ($K^*\bar K$ component for the moment), can be reduced to a compact formula shown in Ref.~\cite{Agatao:2025ckp}
\begin{equation}\label{eq:15}
	T_{K^*\bar K}^{\,\mathrm{tot}}=\frac{
		\tilde{t}_1+\tilde{t}_2+\left(2\, G_0-G_c^{(1)}-G_c^{(2)}\right) \tilde{t}_1 \,\tilde{t}_2}
	{1-G_c^{(1)} \,\tilde{t}_1-G_c^{(2)} \,\tilde{t}_2-\left(G_0^2-G_c^{(1)} \,G_c^{(2)}\right) \tilde{t}_1 \,\tilde{t}_2}.
\end{equation}
The amplitudes $t_1,\,t_2$ are evaluated in the Appendix.

In order to take into account the $\bar K^* K$ component of the $f_1(1285)$, we note that the transition from $K^*$ to $\bar K^*$ requires at least two steps involving the exchange of pseudoscalar mesons and is therefore highly suppressed~\cite{Dias:2021upl}.
Consequently, the total amplitude is
\color{black}
\begin{equation}\label{eq:16}
	(T^{\,\rm tot})^{-1}=\frac{1}{2}\Big(
	(T^{\,\mathrm{tot}}_{K^*\bar K})^{-1}+(T^{\,\mathrm{tot}}_{\bar K^* K})^{-1}
	\Big),
\end{equation}
\color{black}  
\noindent where $T^{\,\mathrm{tot}}_{\bar K^* K}$ is evaluated as $T^{\,\mathrm{tot}}_{K^* \bar K}$ done above replacing $K^*\to \bar K^*;\,\bar K\to  K $. These amplitudes are further discussed in the Appendix.
%
\R{
In Eq.~\eqref{eq:16} we average the inverse of the amplitudes to preserve elastic unitarity. In practice there is no much difference if one uses the average of the amplitudes, but preserving the elastic unitarity is one of the priorities in the present work.
}

\subsection{Scattering length and effective range}
Our amplitude $T^{\,\mathrm{tot}}$ can be related to the standard one of Quantum Mechanics via
\begin{equation}
	\begin{aligned}[b]
		-8\pi\sqrt{s} \left(T^{\,\mathrm{tot}}\right)^{-1} &= (f^{\mathrm{QM}})^{-1} \\[1mm]
		&\approx -\frac{1}{a} + \frac{1}{2}\,r_0 \,q_{\mathrm{cm}}^2 - i \,q_{\mathrm{cm}},
	\end{aligned}
	\label{eq:17}
\end{equation}
with $q_{\mathrm{cm}}$ the $K^+$ momentum in the $K^+f_1(1285)$ rest frame, with $-i\, q_{\mathrm{cm}}$ standing for elastic unitarity. 
As shown in Ref.~\cite{Agatao:2025ckp}, the amplitude $T^{\,\mathrm{tot}}$ satisfies exactly elastic unitarity and we can write
\begin{align}
	a &= \textcolor{black}{\frac{T^{\,\mathrm{tot}}~}{8\pi\sqrt{s}}\Big|_{\mathrm{th}}}, \label{eq:18} \\[2mm]
	r_0 &= \frac{1}{\mu} \left[ \frac{\partial}{\partial \sqrt{s}} \left( -8\pi\sqrt{s} \left(T^{\,\mathrm{tot}}\right)^{-1} + i\,q_{\mathrm{cm}} \right) \right]_{\mathrm{th}}, 
    \label{eq:19}
\end{align}
with $\mu$ the $K^+f_1(1285)$ reduced mass.

{\color{black}{
We find it useful to clarify the concept of elastic unitarity. Even when we have many coupled channels, with some of them open at the threshold of $K^+f_1(1285)$, which render the $K^+f_1$ amplitude complex at threshold, one can still define a complex optical potential 
$V(\sqrt{s})$, such that 
\begin{equation}\label{topt}
	\begin{aligned}[b]
		T=\frac{V_\text{opt}}{1-V_\text{opt}\,G_{Kf_1}};\qquad T^{-1}=V_\text{opt}^{-1}-G_{Kf_1}.
	\end{aligned}
\end{equation}
Taking into account that around the $K^+f_1$ threshold one has
\begin{equation}
	\begin{aligned}[b]
		\sqrt{s} = m_{K^+}+m_{f_1}+\frac{q_{\mathrm{cm}}^2}{2\mu} ,
	\end{aligned}
\end{equation}
one can expand $T^{-1}$ above the $K^+f_1$ threshold in powers of $q_\text{cm}$. Since $V_\text{opt}$ is a function of $\sqrt s$, the expansion of  $V_\text{opt}^{-1}$ gives powers of $q_\text{cm}^2$. So does the expansion of $\text{Re}~G_{Kf_1}$. However,  $\text{Im}~G_{Kf_1}$, which comes when in the propagation of ${K^+f_1}$ in intermediate states the two particles are placed on shell, is linear in $q_\text{cm}$. Actually, from Eqs.~(\ref{eq:17}) and (\ref{topt})  one has
\begin{equation}
	\begin{aligned}[b]
		-8\pi\sqrt{s}\,(-i) \,\text{Im}\,G_{Kf_1}&= 8\pi\sqrt{s} \frac{-i}{8\pi\sqrt{s}}\,q_{\mathrm{cm}} = - i \,q_{\mathrm{cm}}.
	\end{aligned}
	\label{liearterm_eq}
\end{equation}
This means that $a$, $r_0$ can be complex, but the linear term in $q_\text{cm}$ in Eq.(~\ref{eq:17}) is always  $-i~q_\text{cm}$, coming from the elastic propagation of the initial-final state in the intermediate steps of the Lippmann-Schwinger series.
}}

\subsection{Correlation function}
The $K^+f_1(1285)$ correlation function is then written as
\begin{equation}
	\begin{aligned}[b]
		C_{Kf_1}(p) = & 1 + 4\pi \int_{0}^{\infty} \mathrm{d}r \, r^2 \,S_{12}(r) \\[1.5mm]
		& ~~~~~\times \left\{ \left|j_0(pr) + TG\right|^2 - j_0^2(pr) \right\},
	\end{aligned}
	\label{eq:20}
\end{equation}
where{\color{black}{~\cite{Encarnacion:2026zas}

\begin{equation}
	TG = T^\text{tot} \frac{1}{2} \Bigg( G_1(\sqrt{s},r) + G_2(\sqrt{s},r) \Bigg),
	\label{eq:21}
\end{equation}
}}
and $S_{12}(r)$ is the source function
\begin{equation}
	S_{12}(r)=\frac{1}{\left(
		4\pi R^2
		\right)^{3/2}}\;e^{-r^2/4R^2},
	\label{eq:22}
\end{equation}
and $R$ is the radius of the source, with ($i=1,\,2$)
\begin{equation}
		\begin{aligned}[b]
		G_i(\sqrt{s},r) &= \int \frac{\mathrm{d}^3 q}{(2\pi)^3} \,
		\frac{ j_0(qr)\,F_c^{(i)}(q) }{\sqrt{s} - \omega_{K}(\vec q\,) - \omega_c(\vec q\,) + i\epsilon} \\[1.5mm]
		&\quad \times \dfrac{1}{2\,\omega_{K}(\vec q\,)}\, \dfrac{1}{2\,\omega_{c}(\vec q\,)}\;\Theta\left(q_{\mathrm{max}}^{(i)} - q_i^*\right).
	\end{aligned}
	\label{eq:23}
\end{equation}

\section{Results}\label{sec:res}
\begin{figure}[t]
	\begin{center}
		\includegraphics[width=0.48\textwidth]{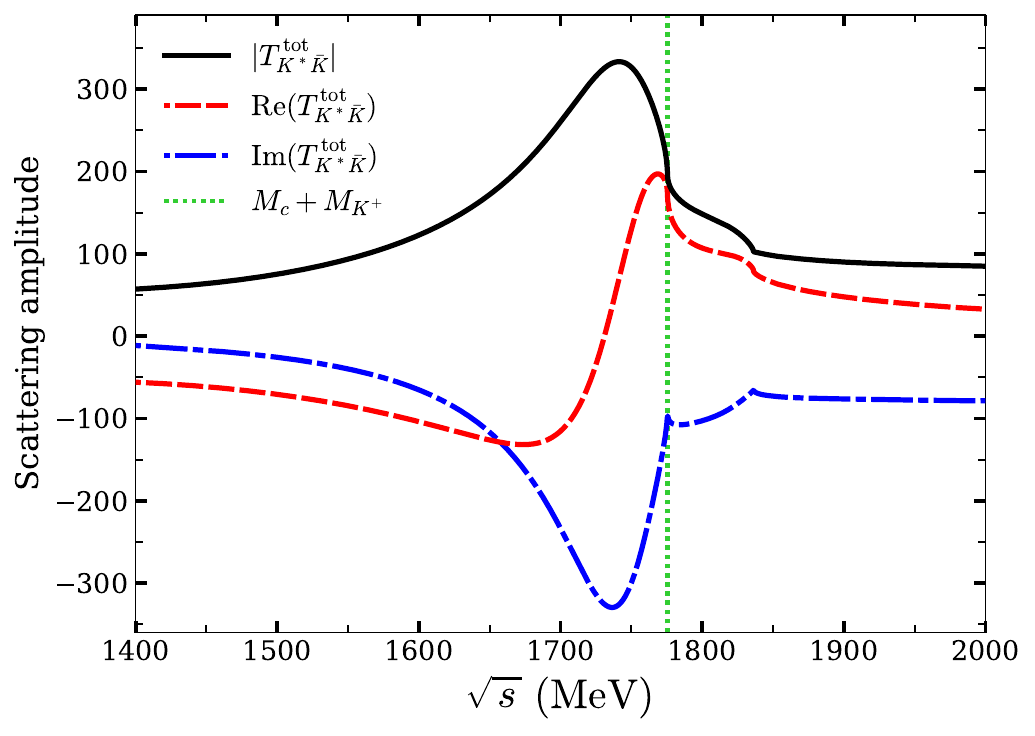}
	\end{center}
	\vspace{-0.5cm}
	\caption{Results for $T_{K^*\bar K}^{\,\mathrm{tot}}$ from Eq.~\eqref{eq:1}. The dashed line represents the real part, the dashed dotted line the imaginary part and the solid line the modulus of the scattering amplitude. 
	The vertical line corresponds to the threshold mass, $M_{K^+}+M_c$.}
	\label{Fig3}
\end{figure}

We begin by presenting the results for the scattering length and effective range obtained from Eqs.~\eqref{eq:18} and \eqref{eq:19}.
We find
{\color{black}
\begin{align}
	a &= (0.53-i\, 0.11)\,\mathrm{fm}, \label{eq:24} \\[2mm]
	r_0 &= (-0.15+i\, 0.29)\,\mathrm{fm}. \label{eq:25}
\end{align}
}
Next we show the results for the three-body scattering amplitude $T_{K^*\bar K}^{\,\mathrm{tot}}$ for the $K^*\bar K$ of Eq.~\eqref{eq:15} in Fig.~\ref{Fig3}.

One can see that the amplitude exhibits an approximately resonant shape, where the magnitude of the imaginary part reaches a peak and the real part goes through zero at the peak of $\left|\mathrm{Im}\,T_{ K^*\bar K}^{\,\mathrm{tot}} \right| $.
This structure suggest that one should expect a bound state close to threshold. However, we still have to consider the second component of Eq.~\eqref{eq:16}, $T_{\bar K^* K}^{\,\mathrm{tot}}$. 
\begin{figure}[t]
	\begin{center}
		\includegraphics[width=0.48\textwidth]{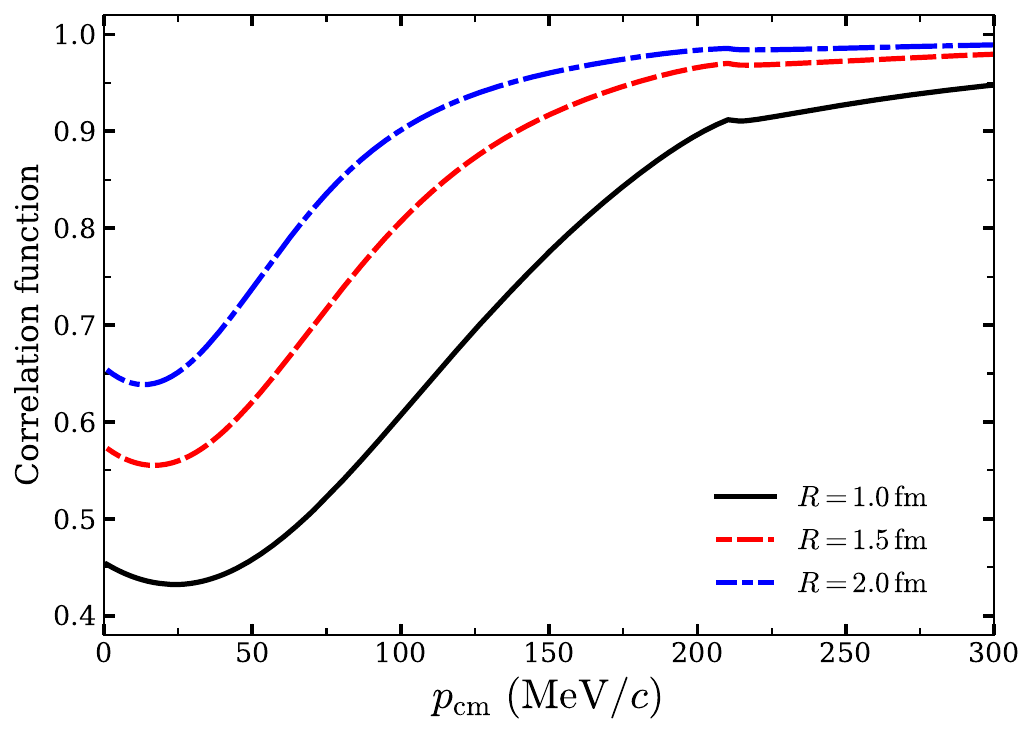}
	\end{center}  
	\vspace{-0.5cm}
	\caption{Correlation function of the component $K^*\bar K$ of the $f_1(1285)$ wave function.}
	\label{Fig4}
\end{figure}
Before presenting these results, we first show in Fig. ~\ref{Fig4}  the correlation function that we obtain from the first component of the cluster, the $K^* \bar K $ component.
We obtain a pattern which is typical of an attractive potential leading to a bound state of the system, consistent with the findings of Ref.~\cite{Encarnacion:2025lyf} in the study of the $pf_1(1285) $ interaction, where also a bound state of the three-body system was found.

Next, we present our final results for the $T^{\, \rm tot}$ and the correlation function, when the two components of the $f_1(1285)$ wave function $K^* \bar K $ and $\bar K^* K $ are considered.
\R{
A detail in Figs.~\ref{Fig3} and \ref{Fig4} deserves a comment. One observes a small kink around $\sqrt s =1830$~MeV in Fig.~\ref{Fig3}, and around $p=210$~MeV in Fig.~\ref{Fig4}， which corresponds to the same  $\sqrt s $. It is easy to see, using Eq.~(\ref{eq:12}), that this corresponds to having $\sqrt s_2 =986$~MeV, which corresponds to the opening of the $K\bar K$ threshold in the amplitude $t_2$. 
}

\begin{figure}[t]
	\begin{center}
		\includegraphics[width=0.48\textwidth]{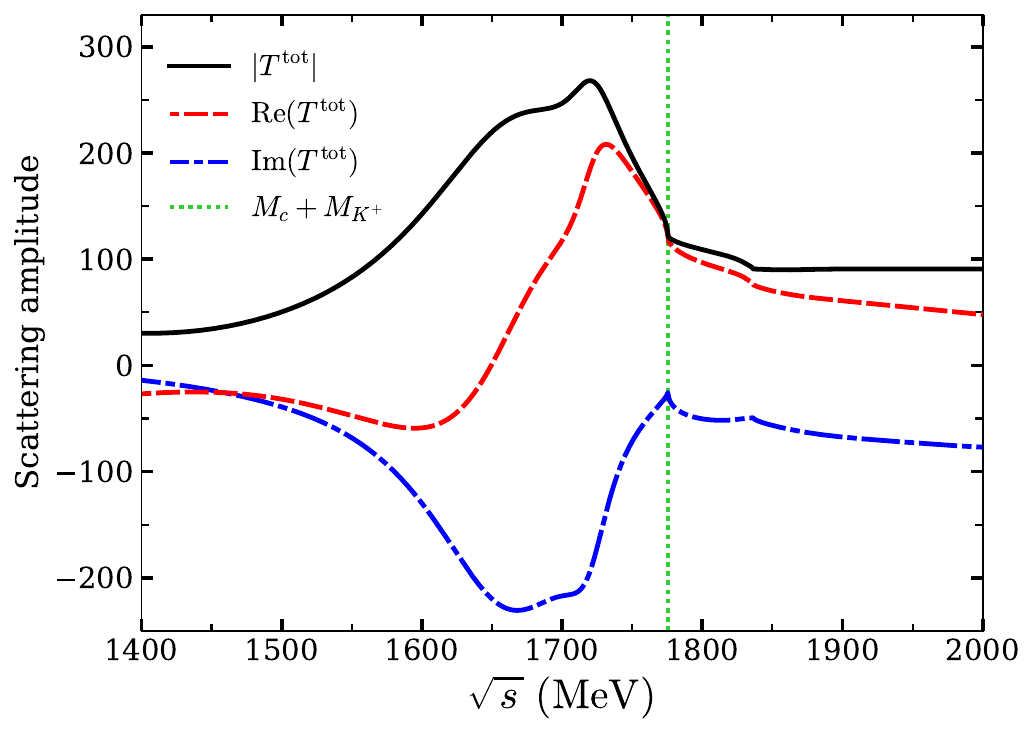}
	\end{center}
	\vspace{-0.5cm}
	\caption{Result for $T^{\, \rm tot}$ of Eq.~\eqref{eq:16}, considering the two components $K^* \bar K$ and $\bar K^* K$ of the $f_1(1285)$ wave function.}
	\label{Fig5}
\end{figure}

Figure~\ref{Fig5} shows the results for $T^{\, \rm tot} $ of Eq.~\eqref{eq:16}.
The results are similar to those obtained in Fig.~\ref{Fig3} for the $K^*\bar K $ component of the wave function. This confirms the resonance-like structure below threshold, which would correspond to a binding with respect to $M_{K^+}+M_c$ of about \R{$56\mev$} and a width of about \R{$123\mev$}.
The structure is clear and stable and deserves some experimental search.
Since it originates from $K K^*\bar K$, and is only weakly bound, it can be investigated in the $KK\pi \bar K $ decay mode, which has a smaller mass ($1619\mev$) than the predicted mass of the state, approximately \R{$1720\mev$}.
Alternatively, one could search for a kaon in coincidence with one of the decay modes of $f_1(1285)$. 
Yet, all significant decay modes of the $f_1(1285)$ contain three final particles, with the most important one, $a_0(980)\pi$, decaying further into $\pi\pi\eta$.
Certainly, observing the structure in different decay modes would provide further evidence for the existence for this state.
Reconstructing invariant masses of four particles is routine in the LHCb collaboration, for instance to detect $B$ mesons in the $\bar B^0\to D^+ \pi^-\pi^-\pi^+$ decay or $D^0$ in the $D^0\to K^-\pi^+\pi^-\pi^+$ decay \cite{LHCb:2023yjo,LHCb:2012ssw}.
It is also done by the ALICE collaboration identifying $pK\bar K \pi$ in their measurement of the correlation function of the $p f_1(1285)$ \cite{Privatecomm}.

\begin{figure}[t]
	\begin{center}
		\includegraphics[width=0.48\textwidth]{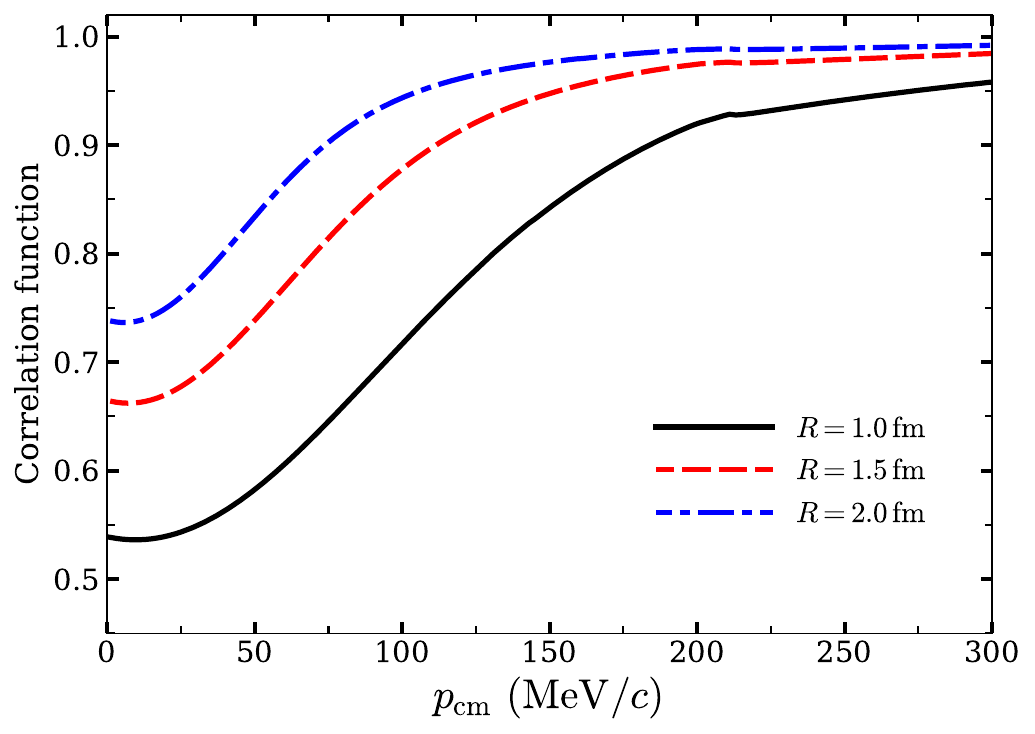}
	\end{center}
	\vspace{-0.5cm}
	\caption{Correlation function considering the two components $K^* \bar K$ and $\bar K^* K$ of the $f_1$ wave function.}
	\label{Fig6}
\end{figure}

Furthermore, we show the correlation function corresponding to the two components $K^*\bar K $ and $\bar K^* K $ of the $f_1(1285)$ in Fig.~\ref{Fig6}.

We conclude with one observation.
In Ref.~\cite{Yan:2023vbh}, the interaction of pseudoscalar mesons with axial vector mesons is studied using the Weinberg-Tomozawa (WT) term derived from chiral Lagrangians at the leading order of the chiral expansion by treating the axial vector mesons as matter ﬁelds and the pseudoscalar mesons as the pseudo-Nambu-Goldstone bosons resulting from the spontaneous breaking of chiral symmetry.  
This procedure provides the $1/f^2_\pi$ terms of the interaction, the leading-order potential $V$. 
As shown in Table 8 of that paper, the $K f_1(1285)$ potential is identically zero. In contrast, assuming the $f_1(1285)$ to be a $K \bar K^*, \bar K K^*$ molecular state, we have found a sizeable strength which is suﬀicient to produce a bound state, and a correlation function significantly different from unity.
It is interesting to note that our combination of amplitudes of Eq. \eqref{eq:2} for $KK$ and $K \bar K$ interaction, together with Eq. \eqref{eq:A3}, which ignores the identity of two $K^+$, and Eq. (18) of Ref. \cite{Oller:1997ti}, are opposite to each other (the same can be said about the $K K^*$ and $K \bar K^*$, which now are not identical particles). 
This means that replacing $t_2$ by $V_2$, the $V_2$ term for $K \bar K$ and the one for $K K$ would cancel in the sum of our Eq.~\eqref{eq:16} at order of $1/f_\pi ^2$.  
However, as we have seen, considering the $f_1(1285)$ as a bound state, the amplitude $T^{\rm tot}$ for $K f_1(1285)$, prior to the elastic unitarization, is given by Eq.~\eqref{eq:15}, and one starts from $t_1, t_2$, instead of $V_1, V_2$, where the unitarization of the $K \bar K$, or $KK$ amplitudes is already implemented, going from $V$ to $t$. 
This is a reminder of the $t \rho$ optical potential, with $\rho$ the nuclear density, that one obtains when studying the interaction of an external particle with a nucleus. 
Then, the $K \bar K$ interaction is attractive while the $KK$ one is repulsive, and in this case the $t$ matrices are very different, because the attractive potential gives an amplitude containing a pole, while the repulsive potential gives rise to a smooth structure and there is no cancellation.  
In addition, as a consequence of the molecular picture for the $f_1(1285)$ state, we have the sums of terms for Figs. \ref{Fig1} and \ref{Fig2} that lead to the final formulas in Eqs. \eqref{eq:15} and \eqref{eq:16}.

{\color{black}{
\section{Results with convolution to consider the width of the $f_1(1285)$}

So far all the results are obtained using the nominal mass of the $f_1(1285)$, ignoring its width, $\Gamma_{f_1} \sim 23$~MeV. The consideration of the $f_1(1285)$ width should smear the results around threshold and we show here the results in this case. For this we take the magnitudes, $a$ of  Eq.~(\ref{eq:18}), $r_0$ of  Eq.~(\ref{eq:19}) and $(T^\text{tot})^{-1}$ of Eq.~(\ref{eq:16}). These magnitudes depend on the mass of the $f_1$ and then we make a convolution of these results with the spectral function of the $f_1(1285)$. 
We take
\begin{align}
   \Bigg( \tilde{T}^\text{tot} \Bigg)^{-1}= \frac{1}{C}&\int_{(m_{f_1}-2\Gamma_{f_1})^2}^{(m_{f_1}+2\Gamma_{f_1})^2} d\tilde m^2  \Bigg( {T}^\text{tot}(\tilde m)\Bigg)^{-1} \notag\\
    & \times \left(-\frac{1}{\pi}\right) \text{Im} \left(\frac{1}{\tilde m^2-m_{f_1}^2+im_{f_1}\Gamma_{f_1}}\right),
\end{align}
%
%
\begin{align}
    C=  \int_{(m_{f_1}-2\Gamma_{f_1})^2}^{(m_{f_1}+2\Gamma_{f_1})^2}  d\tilde m^2 (-\frac{1}{\pi}) \text{Im} (\frac{1}{\tilde m^2-m_{f_1}^2+im_{f_1}\Gamma_{f_1}}),
\end{align}
and then from $\tilde{T}^\text{tot}$  we evaluate the observables.
The new results for $a,~r_0$ are 
\begin{align}
	a &= (0.55 - i\, 0.20)\,\mathrm{fm}, \label{eq:arconvo_a} \\[2mm]
	r_0 &= (-1.85 + i \,0.26)\,\mathrm{fm}. \label{eq:arconvo_r}
\end{align}
Compared with the results of Eqs.~(\ref{eq:24}), (\ref{eq:25}) we observe a moderate change in $a$, but a bigger one in $r_0$.

In Fig.~\ref{Fig1new} we show the results for the scattering matrix with and without the convolution. Compared with the results of Fig.~\ref{Fig5} the changes are small, affecting mostly the region around threshold. The structure for the bound state is unaltered.
\begin{figure}[!t]
	\begin{center}
		\includegraphics[width=0.48\textwidth]{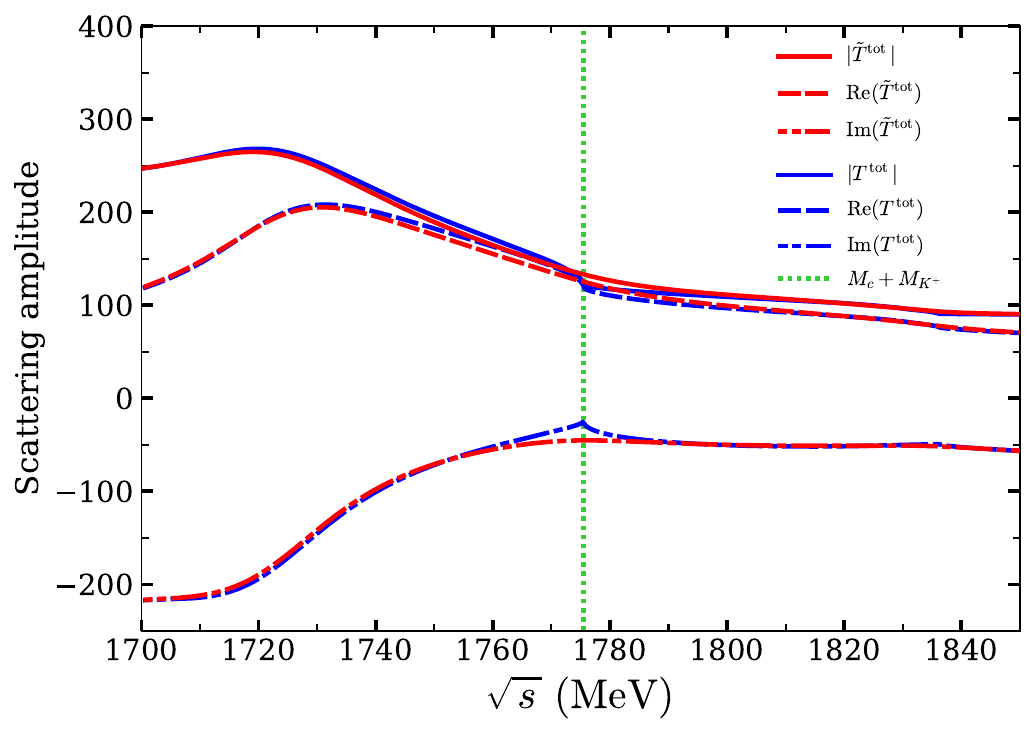}
	\end{center}
	\vspace{-0.5cm}
	\caption{Comparison between the amplitude $\tilde{T}^{\text{tot}}$ with convolution (red lines) and the amplitude $T^{\text{tot}}$ without convolution (blue lines).}
	\label{Fig1new}
\end{figure}
In Fig.~\ref{Fig2new} we show the results for the correlation functions with  and without the convolution.
\begin{figure}[h]
	\begin{center}
		\includegraphics[width=0.48\textwidth]{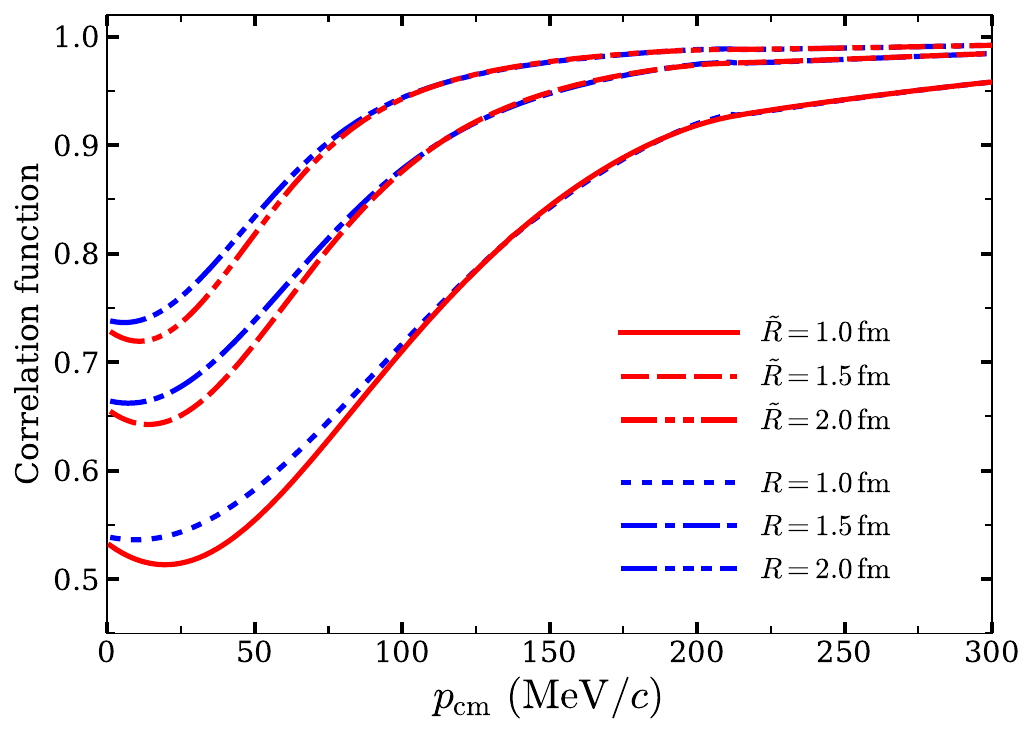}
	\end{center}
	\vspace{-0.5cm}
	\caption{Comparison between the correlation functions with and without convolution. The results with tilde (red lines) are for convolution.}
	\label{Fig2new}
\end{figure}
As we can seen, the
effects on the correlation functions due to the convolution are relatively small. One finds a small decrease around $\vec{p}=25~\text{MeV}/c$ and no change at high momenta.
}
}

\section {Conclusions}
We have investigated the interaction of a kaon with the $f_1(1285)$ resonance, evaluating the three-body amplitude, the scattering length and effective range of the $K f_1$ system, and the corresponding correlation function. 
The motivation is the new wave of experiments undertaken by the ALICE collaboration, starting by the measurement of the $p f_1(1285)$ correlation function. 
We use a framework based on the fixed center approximation (FCA), in which the cluster is the $f_1(1285)$ and the external particle is a kaon, with the $f_1(1285)$ being a molecule made of $K^* \bar K$ and $\bar K^* K$. 
The framework of the FCA is improved to implement exact elastic unitarity, which is necessary to evaluate the scattering parameters and the correlation function. 
  
We find that the $K f_1$ amplitude develops a resonant structure about \R{$56\mev$} below threshold, with a width of about \R{$123\mev$}. 
We discuss that this structure requires the measurement of the invariant mass of four particles, but this is a technique used both by the ALICE and LHCb collaborations. 
  
The correlation function has a structure corresponding to having a bound state below the $K f_1$ threshold, or resonant state, since it has a width, and eventually the measure of this correlation function could already contain sufficient information to predict the existence of this bound state, using the inverse method, successfully tested in different contexts \cite{Ikeno:2023ojl,Albaladejo:2023wmv}. 
  
We also discuss the difference found for the interaction of the $K$ with the $f_1(1285)$ from the molecular picture of the $f_1$, compared with that obtained assuming the $f_1$ as an \R{ordinary, non-molecular,} particle. 
We obtain a strong attractive interaction, sufficient to produce a bound state, compared to the case where the $f_1$ is an \R{ordinary} particle where the interaction is found to be zero.
   
All these reasons should motivate the measurement of the correlation function and searches for the four body invariant mass distributions to learn more about the nature of the $f_1(1285)$ resonance, and related axial vector meson resonances by extension, and the possible existence of a $K f_1(1285)$ bound state.

{\color{black}{
Note: After this paper was submitted for publication, important developments concerning the present paper occurred. The experimental work on the $p f_1(1285)$ correlation function of Ref.~\cite{Privatecomm} was finalized and presented in the Strangeness and Quark Matter Conference~\cite{Serkšnytė:2026}.  The results were compared with the updated theoretical results of Ref.~\cite{Encarnacion:2025lyf} given in Ref.~\cite{Encarnacion:2026zas}, and a good agreement was found in the shape and strength of the correlation function. These results certainly provide support for the structure assumed here for the $f_1(1285)$ and the method used to deal with the interaction of an external particle with the $f_1(1285)$ resonance.

The data of Ref.~\cite{Serkšnytė:2026} also show that there are no Coulomb effects close to the $pf_1$ threshold, as it should be since the $f_1$ is neutral. There could be Coulomb effects in the amplitudes $t_1, ~t_2$, which we do not consider. They should be smeared in the correlation function, but in our case they should appear in the thresholds of $K^+K$ and $K^+K^*$ which,  using Eq.~(\ref{eq:12}), correspond to $p=210$~MeV$/c$ and 232~MeV$/c$, respectively,  a region far away from the relevant one for the correlation function  where one sees strong deviations from unity, or the scattering amplitude below threshold  where the bound state appears.

The experimental results from Ref.~\cite{Serkšnytė:2026} are also most welcome in relation to an issue that has sparked some discussions. The first one has to do with the fact that the source function and the correlation function are not invariant under unitary transformations~\cite{Epelbaum:2025aan}. While this is certainly true, it was found in Ref.~\cite{Gobel:2025afq} and  Ref.~\cite{Molina:2025lzw} that it has no practical consequences in realistic calculations. There are other potential problems in the definition and construction of the correlation functions, as discussed in Ref.~\cite{Albaladejo:2025lhn}. Yet, the answer to these potential problems has been provided by the experimental results of Ref.~\cite{Serkšnytė:2026}. The agreement of the experiment data for the $pf_1(1285)$ correlation function with the predictions made in Refs.~\cite{Encarnacion:2025lyf,Encarnacion:2026zas}, using the formalism of the present paper, and the idea of the $f_1(1285)$ as a molecular state, strongly support the accuracy of the predictions made in the present work.
}}

\section*{Acknowledgments}
We would like to thank Prof. Juan Nieves for useful discussions.
This work is partly supported by the National Natural Science Foundation of China (NSFC) under Grants No. 12575081, No. 12365019, No. 12405089 and No. 12247108,
and by the Natural Science Foundation of Guangxi province under Grant No. 2023JJA110076,
and by the Central Government Guidance Funds for Local Scientific and Technological Development, China (No. Guike ZY22096024).
This work is partly supported by the China Postdoctoral Science Foundation under Grant
No. 2022M720360 and No. 2022M720359. 
 J. Song would also like to thank the support from the Hainan Provincial Excellent Talent Team under the “Four Talents” Gathering Program of Hainan Province.
This work was supported by the National Key R{\&}D Program of China (Grant No. 2024YFE0105200).
This work is also partly supported by the Spanish Ministerio de Economia y Competitividad (MINECO) and European FEDER funds under Contracts No. FIS2017-84038-C2-1-PB, PID2020-112777GB-I00, and by Generalitat Valenciana under contract PROMETEO/2020/023. 
This project has received funding from the European Union Horizon 2020 research and innovation program under the program H2020-INFRAIA-2018-1, grant agreement No. 824093 of the STRONG-2020 project.

\appendix
\section{Amplitudes $t_1,\,t_2$}
All the amplitudes are calculated using coupled channels and the interaction driven by the exchange of vector mesons following the local hidden gauge approach \cite{Bando:1984ej,Bando:1987br,Meissner:1987ge,Nagahiro:2008cv}. 
In the case of pseudoscalar-vector interactions, $K\bar K^*,\,\bar K K^* $, the amplitudes are taken from Ref.~\cite{Roca:2005nm}.

The $KK,\,K\bar K$ interactions are obtained using the Lagrangian 
\begin{equation}
	\mathcal{L}_{VPP}=-\mathrm{i}g\left\langle \left[ P,\,\partial_{\mu}P\right]V^{\mu}\right\rangle,
	\label{eq:A1}
\end{equation}
with $g=\frac{M_V}{2f},\,M_V=800\mev$ and $f=93\mev$.

\subsection{The $KK$ interaction}

1) $KK$ interaction in $I=1$\\

\begin{figure}[b]
	\begin{center}
		\includegraphics[width=0.46\textwidth]{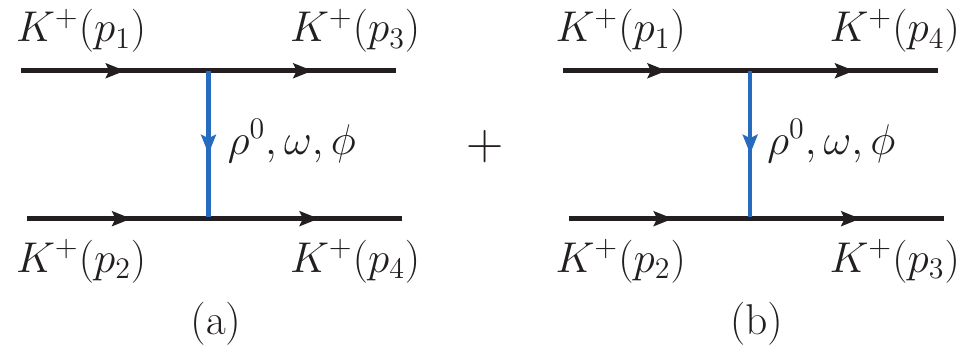}
	\end{center}
	\vspace{-0.5cm}
	\caption{Diagrams for the $K^+K^+\to K^+K^+$ interaction: (a) direct diagram, (b) crossed diagram.}
	\label{FigA1}
\end{figure}

We take $K^+K^+\to K^+K^+$ and the potential is obtained according to Fig.~\ref{FigA1}.
Note that due to the identity of the particles we must consider symmetrized amplitudes and we must take diagrams (a) and (b) of Fig.~\ref{FigA1}. 
From the diagram (a), summing the contributions of $\rho^0,\,\omega,\,\phi$ exchange, we obtain
\begin{equation}
	V^{(a)}=\frac{2g^2}{M_V^2}\left(p_1+p_3\right)\left(p_2+p_4\right),
	\label{eq:A2}
\end{equation}
which projected in $S$-wave leads to
\begin{equation}
	\begin{aligned}[b]
		V^{(a)}&=\frac{2g^2}{M_V^2}\left[
		\frac{3}{2}s-\frac{1}{2}\left(
		m_1^2+m_2^2+m_3^2+m_4^2
		\right) \right.\\
		&\quad \left. -\frac{1}{2s}\left(
		m_1^2-m_2^2
		\right)\left(
		m_3^2-m_4^2
		\right) \right].
	\end{aligned}
	\label{eq:A3}
\end{equation}
Summing the contribution of diagram (b) we obtain
\begin{equation}
	V_{KK}^{I=1}=\frac{4g^2}{M_V^2}\left(
	\frac{3}{2}s-2M_K^2
	\right).
	\label{eq:A4}
\end{equation}
The $T$ matrix for this single channel case is given by
\begin{equation}
	T=\frac{V}{1-\frac{1}{2}VG}.
	\label{eq:A5}
\end{equation}
The presence of $\frac{1}{2}G$, instead of the ordinary $G$ function in Eq.~\eqref{eq:A5} is due to the identity of the two $K^+$.

The $G$ function is taken from Ref.~\cite{Lin:2021isc} using the cutoff method and is regularized with a maximum $\left|\vec{q}\,\right|$ value, $q_{\rm max} = 650\mev$.
\R{
This would correspond to $q_{\rm max}^\text{(2)}$ for the second block $\bar K^*K$ of the $f_1$ wave function.
}

2) $KK$ interaction in $I=0$ \\

The $I=0$ $KK$ wave function is
\begin{equation}
	| KK,\,I=0\rangle=\frac{1}{\sqrt{2}}\left(K^+K^0-K^0K^+\right),
\end{equation} 
and in $S$-wave $K^+K^0$ and $K^0K^+$ are the same and the interaction vanishes. Hence
\begin{equation}
	V_{KK}^{I=0}=0.
	\label{eq:A6}
\end{equation}

\subsection{The $KK^*$ interaction}

1) $KK^*$ interaction in $I=1$\\

We need now an extra Lagrangian for the $VVV$ vertex given in the limit of small three momenta by \cite{Bando:1984ej,Bando:1987br,Meissner:1987ge,Nagahiro:2008cv}
\begin{equation}
	\mathcal{L}_{VVV}=i g\;\langle \,
	 \left(
	  V^{\mu}\partial_{\nu}V_{\mu}-\partial_{\nu}V^{\mu}V_{\mu} \right)V^{\nu} \,\rangle.
	\label{eq:A7}
\end{equation}
In this limit $V^{\nu}$ must be the exchanged vector, because otherwise $\epsilon^{\nu}$ has only the spatial components and $\partial_{\nu}$ becomes a three momentum which is zero.
Then the interaction is the same as for $KK$ in $I=1$ except for an $\vec \epsilon\; \vec \epsilon'$ factor for the external vectors which factorizes in all amplitudes and plays no role in the binding and shapes of the amplitudes.
Note that the diagram equivalent to Fig.~\ref{FigA1}(b) has now the $K\to K^*$ transition driven by $\pi$ exchange, which was found very small compared to vector exchange in Ref.~\cite{Dias:2021upl} and we ignore it here.

Then the $KK^*,\,I=1$ scattering matrix is given by
\begin{equation}
	T=\frac{V'}{1-V'G\;},
	\label{eq:A8}
\end{equation}
where
\begin{equation}
	V'=\frac{2g^2}{M_V^2}\left[
		\frac{3}{2}s-(
		M_K^2+M_{K^*}^2)
		-\frac{1}{2s}(
		M_{K^*}^2-M_K^2) \right],
		\label{eq:A9}
\end{equation}
and $G$ stands now for $KK^*$ propagation. Note that we do not have now the $\frac{1}{2}G$ in Eq.~\eqref{eq:A8} since $K,\,K^*$ are not identical particles.\\

\R{
To take into account the $K^*$ width we perform a convolution of the $G$-function, as done in the Appendix of Ref.~\cite{Agatao:2025ckp}. We use cutoff regularization for the $G$-function, using a value of    $q_\text{max}=1000$~MeV, as in the case of the $K\bar K^*$ interaction, which would correspond to $q_\text{max}^\text{(1)}$  for the $K^*\bar K$ block of the $f_1$ wave function.
}

2) $KK^*$ interaction in $I=0$\\

Here 
\begin{equation}
	| K^*K,\,I=0 \rangle=\frac{1}{\sqrt{2}} \Big( K^{*+}K^0-K^{*0}K^+ \Big).
\end{equation}
We cannot use here the argument that $K^{*+}K^0$ is the same as $K^{*0}K^+$ in $S$-wave, used for the case of $KK$, but since the vertices involved have been shown to be equivalent for $KKV$ and $K^*K^*V$, then $V_{KK^*}^{I=0}=0$, as can also be obtained explicitly taking $m_{\omega}=m_{\rho^0}$.

\subsection{The $K\bar K$ amplitudes}

Here we take these amplitudes from the work of Ref.~\cite{Lin:2021isc}, which are evaluated using the $\pi^+\pi^-$, $\pi^0\pi^0$, $K^+K^-$, $K^0K^0$, $\pi^0\eta$, $\eta\eta$ channels, updating the work of Ref.~\cite{Oller:1997ti} to account explicitly for the $\eta-\eta'$ mixing of Ref.~\cite{Bramon:1992kr}.
With the usual isospin phase convention $(K^+,\,K^0)$, $(\bar K^0,\,-K^-)$, we have 
\begin{equation}
	\begin{aligned}
		| K\bar K,\,I=0\rangle &= -\frac{1}{\sqrt{2}} \Big(K^{+}K^- + K^{0}\bar K^0 \Big),\\[2mm]
		| K\bar K, \,I=1, \,I_3=0 \rangle &= -\frac{1}{\sqrt{2}}\Big( K^{+}K^- - K^{0}\bar K^0\Big),
	\end{aligned}
	\label{eq:11}
\end{equation}
and then
\begin{equation}
	\begin{split}
		&\langle 
		K\bar K, \,I=0 \left|\, t \,\right|K\bar K,\,I=0 \rangle\\[1mm]
		=&\frac{1}{2}\Big(
		t_{K^+K^-,\,K^+K^-}+2\,t_{K^+K^-,\,K^0\bar K^0}+t_{K^0\bar K^0,\,K^0\bar K^0}
		\Big),\\[2.5mm]
		&\langle K\bar K, \,I=1 |\, t \,| K\bar K, \,I=1
		\rangle\\[1mm]
		=&\frac{1}{2}\Big(
		t_{K^+K^-,\,K^+K^-}-2\,t_{K^+K^-,\,K^0\bar K^0}+t_{K^0\bar K^0,\,K^0\bar K^0}
		\Big).
	\end{split}
	\label{eq:A10}
\end{equation}
\R{
In Ref.~\cite{Lin:2021isc}, $q_\text{max}$ corresponding to $q_\text{max}^\text{(2)}$ in the $K^*\bar K$ block of the $f_1$ wave function, is taken as $q_\text{max}=650$~MeV.
}

\subsection{The $K\bar K^*$ amplitudes}

Here we take advantage that these amplitudes are evaluated in Ref.~\cite{Roca:2005nm} and we use information obtained in that work. 
In that work the interaction leads to the dynamical generation of the different axial vector resonances, which are classified by spin and $C$-parity as shown in Table~\ref{table1}.
\begin{table}
	\caption{Classification of the axial vector resonances. The old $h_1(1380)$ has been updated in the PDG to the $h_1(1415)$.}
	\label{table1}
	\setlength{\tabcolsep}{11pt}
	\begin{spacing}{2}
		\begin{tabular}[c]{ccccc}
			\hline\hline
			$J^{PC}$ & $I=1$ & $I=0$  \\
			\hline
			$1^{+-}$ & $b_1(1235)$ & $h_1(1170),\,h_1(1380)\,[h_1(1415)]$\\
			$1^{++}$ & $a_1(1260)$ & $f_1(1285)$\\
			\hline\hline
		\end{tabular}
	\end{spacing}
\end{table}

Note that the $f_1(1420)$ is dismissed as a different resonance in Refs.~\cite{Debastiani:2016xgg,Lin:2023ajb}, where it is shown that it is the manifestation of the $f_1(1285)$ in the $K \bar K^*$ decay mode.

To use the information of Ref.~\cite{Roca:2005nm}, we use the following strategy: We need the $K\bar K^*$ amplitude, but the physical states are combinations of $K\bar K^* $ and $\bar K K^* $.
We use now the isospin multiplets $(K^+,\,K^0)$, $(\bar K^0,\,-K^-)$, $(K^{*+},\,K^{*0})$, $(\bar K^{*0},\,-K^{*-})$, and take into account the $CK=\bar K,\, CK^*=-\bar K^*$. 
Then we have

1) $I=0,\,C=+,\,f_1(1285)$
\begin{equation}
	\begin{aligned}[b]
		f_1(1285)&\equiv \frac{1}{\sqrt{2}}\left[
		-\frac{1}{\sqrt{2}}\left(
		K^+K^{*-} + K^0\bar K^{*0}
		\right) \right.\\
		&\quad \left. + \frac{1}{\sqrt{2}}\left(
		K^-K^{*+} + \bar K^0 K^{*0}
		\right)
		\right] \\
		& =\frac{1}{\sqrt{2}}\left[
		K\bar K^*(I=0) + \bar K K^*(I=0)
		\right].
	\end{aligned}
	\label{eq:A11}
\end{equation}

2) $I=1,\,C=+,\,a_1(1260)$
\begin{equation}
	\begin{aligned}[b]
		a_1(1260)&\equiv \frac{1}{\sqrt{2}}\left[
		-\frac{1}{\sqrt{2}}\left(
		K^+K^{*-} - K^0\bar K^{*0}
		\right) \right.\\
		&\quad \left. + \frac{1}{\sqrt{2}}\left(
		K^-K^{*+} - \bar K^0 K^{*0}
		\right)
		\right] \\
		& =\frac{1}{\sqrt{2}}\left[
		K\bar K^*(I=1) - \bar K K^*(I=1)
		\right].
	\end{aligned}
	\label{eq:A12}
\end{equation}
Hence
\begin{equation}
	\begin{aligned}[b]
		\langle f_1 |\, t \,|f_1 \rangle
		&= \frac{1}{2}\Big[
		\langle 
		K\bar K^*,\,I=0\left|\, t \, \right|K\bar K^*,\,I=0
		\rangle \\[1.5mm]
		&\quad  +2 \, \langle
		K\bar K^*,\,I=0 | \, t \,|\bar K K^*,\,I=0
		\rangle \\[1.5mm]
		&\quad + \langle
		\bar K K^*,\,I=0 |\, t \, |\bar K K^*,\,I=0
		\rangle \Big].
	\end{aligned}
	\label{eq:A13}
\end{equation}

3) $I=0,\,C=-,\,h_1(1170),\,h_1(1415)$\\

\begin{table*}[t]
\color{black}
	\caption{Masses, widths and couplings $g_i$ of the $J^P=1^+$ resonances to the  $K\bar K^*$ component (without the convolution of the $K^*$ width), taken from Ref.~\cite{Roca:2005nm} and PDG masses and widths. [in the unit of MeV].}
	\label{table2}
    \setlength{\tabcolsep}{22pt}
	\begin{spacing}{1.8}
		\begin{tabular}[c]{cccccc}
			\hline\hline
			Resonance & Mass &$\Gamma$  & $g_i $   & Mass~(PDG)&$\Gamma$~(PDG) \\
			\hline
			$h_1(1415)$ &$1245$ &$14$ & $6147+ i\, 183$ &$1400^{+9}_{-8}$ & $78\pm11$ \\
			$f_1(1285)$ &$1288$ &$0$ & $7230$ &$1281.8\pm0.5 $& $23.0\pm1.1$\\
			$b_1(1235)$ &$1247$ &$56$ & $6172-i\, 75$ &$1229.5\pm3.2$ & $142\pm9$\\
			$a_1(1260)$ &$1011$ &$168$ & $1872-i \, 1486$ &$1230\pm40$ & $250-600$\\
			\hline\hline
		\end{tabular}
	\end{spacing}
\end{table*}

Similarly to Eq.~\eqref{eq:A11} we will now have
\begin{equation}
	h_1 \equiv \frac{1}{\sqrt{2}}\Big[
	K\bar K^*(I=0) - \bar K K^*(I=0)
	\Big],
	\label{eq:A14}
\end{equation}
and
\begin{align}\label{eq:A15}
	\langle h_1 |\, t \,| h_1 \rangle
	&=\frac{1}{2}\Big[ \langle K\bar K^*,\,I=0 |\, t \,| K\bar K^*,\,I=0 \rangle 
	\nonumber \\
	&-2\, \langle K\bar K^*, \,I=0 |\, t \,| \bar K K^*, \,I=0 \rangle \nonumber \\
	&+ \langle \bar K K^*, \,I=0 |\, t \,|\bar K K^*, \,I=0 \rangle \Big].
\end{align}	
Hence we find
\begin{align}\label{eq:A16}
	&\langle f_1 |\, t \,|f_1 \rangle + \langle h_1 |\, t \,| h_1 \rangle \nonumber\\[1mm]
		=& \langle K\bar K^*,\,I=0 |\, t \,| K\bar K^*,\,I=0 \rangle \nonumber\\[1mm]
		& + \langle \bar K K^*,\,I=0 |\, t \, |\bar K K^*,\,I=0 \rangle \nonumber\\[1mm]
		=& 2 \langle K\bar K^*,\,I=0 |\, t \, | K\bar K^*,\,I=0 \rangle,
\end{align}
where we have made use of invariance over $C$ parity of the reaction in the last step.\\

4) $I=1,\,C=-,\,b_1(1235)$\\

Here we have analogy to in Eq.~\eqref{eq:A12}
\begin{equation}
	b_1(1235)=\frac{1}{\sqrt{2}}\Big[
	K\bar K^*(I=1)+\bar K K^*(I=1)
	\Big],
	\label{eq:A17}
\end{equation}
and analogously to Eq.~\eqref{eq:A16} we have now
\begin{equation}
	\begin{aligned}[b]
		\left< 
		a_1\right| t\left|a_1
		\right> + \left< 
		b_1\right| t\left|b_1
		\right>
		= 2 \left< 
		K\bar K^*,\,I=1\right| t\left| K\bar K^*,\,I=1
		\right> \,.
	\end{aligned}
	\label{eq:A18}
\end{equation}
By means of Eqs.~\eqref{eq:A16}, \eqref{eq:A18}, we can obtain the amplitudes that we need for $K\bar K^*$ in $I=0$ and $I=1$.

In the case of Eq.~\eqref{eq:A16}, we would have to sum the contributions of the $h_1(1170)$ and $h_1(1415)$. 
However, the coupling of the $h_1(1170)$ to the $K\bar K^*$ component is very small compared to the $h_1(1415)$ ($2\%$ ratio in $g^2$ \cite{Roca:2005nm}), and can be neglected. Hence, the $h_1(1415)$ is the only resonance relevant in the relationship of Eq.~\eqref{eq:A16}.

The next step is to evaluate these amplitudes and then we parametrize them with Breit-Wigner forms as
\begin{equation}
	\left<R_i\right|t\left|R_i\right>=\frac{g_i^2}{s-M_{R_i}^2+i M_{R_i}\Gamma_{R_i}},
	\label{eq:A19}
\end{equation}
taking $M_{R_i},\,\Gamma_{R_i}$ from the PDG~\cite{ParticleDataGroup:2024cfk} and the couplings $g_i$ from Ref.~\cite{Roca:2005nm}.

We should note that the masses obtained in Ref.~\cite{Roca:2005nm} differ somewhat from those of the PDG for the $a_1(1260)$ and the $h_1(1415)$. 
However, the couplings are smoothly dependent on the binding energy and one has \cite{Gamermann:2009uq}
\begin{equation}
	g\sim E_R\left(16\pi \gamma /\mu\right)^{1/2}\,,
	\qquad
	\gamma=\sqrt{2\mu B}\,\,,
	\label{eq:A20}
\end{equation}
with $\mu$ the reduced mass of $KK^*$ and $B$ the binding of the resonance, hence $g$ goes as $B^{1/4}$.
We, thus, multiply the coupling of Ref.~\cite{Roca:2005nm} by 0.76 for the $a_1(1260)$ and 0.51 for the $h_1(1415)$, but we have observed that the effect of these changes is minimal, compared to the use of the original couplings.
The couplings of Ref.~\cite{Roca:2005nm} to $K \bar K^*$ are given in Table~\ref{table2}.
\R{
We also need $q_\text{max}$ in this case to be used in the different equations. In this case, following Ref.~\cite{Roca:2005nm} we take $q_\text{max}=1000$~MeV, corresponding to $q_\text{max}^\text{(1)}$ in the $\bar K^*K$  block of the $f_1$ wave function.
}


\bibliography{refs.bib} 
\end{document}